\documentstyle[preprint,aps]{revtex}
\begin{document}
\draft
\title{Optical response of small carbon clusters}
\author{K. Yabana} 
\address{Department of Physics, Niigata University\\
Niigata, Japan\\
and\\}
\author{G.F. Bertsch\footnote{E-mail bertsch@phys.washington.edu}}
\address{Physics Department and Institute for Nuclear
Theory\\
University of Washington, Seattle, WA 98195 USA
}
\maketitle
\def\C60{C$_{60}$}
\def\eq#1{eq. (\ref{#1})}
\begin{abstract}
We apply the time-dependent local density approximation (TDLDA)
to calculate
dipole excitations in small carbon clusters. A strong low-frequency
mode is found which agrees well with observation for clusters C$_n$
with $n$ in the range $7- 15$.  The size dependence of the
mode may be understood simply as the classical resonance of electrons
in a conducting needle.  For a ring geometry, the lowest collective
mode occurs at about twice the frequency of the collective
mode in the linear chain, and this may also be understood in
simple terms.
\end{abstract}
\pacs{version: 22.8.97}

\def\be{\begin{equation}}
\def\ee{\end{equation}} 

\section{Introduction}  
  In
this work we examine strong optical transitions in light carbon clusters,
with the goal of providing a diagnostic tool for the structure of the
clusters.
Carbon has a rather complex 
structural evolution \cite{we89} as a function of the number of atoms $n$.
Small clusters are linear, and at higher $n$ rings are favored.  For
intermediate $n$ values, rings may be more favored at even $n$ and
linear chains at odd $n$.  Direct information on the shapes of the
clusters may be obtained by ion chromatography \cite{he93}, but one
mostly relies on spectroscopic data to interpret the structure \cite{ya88}.

We will demonstrate that the strong optical absoption resonances behave 
in a very systematic way as function of cluster size, and that the frequency
of the transitions gives direct information about the shape of 
the cluster.  In particular, the strong $\pi-\pi^*$ transition in 
chains is at about
half the frequency of the mode in rings.  We also find that the 
frequencies can be understood very simply using classical concepts.  
The most important determinant in systems of similar composition is the 
polarizability, which is much larger for chains along their axes than for 
rings in the plane of the ring.

In the section 2 below we discuss our quantum calculations, which
are done using the time-dependent local-density approximation (TDLDA).
This is nothing more than the Kohn-Sham equations with the ``energy"
of the orbital replaced by the time derivative $i\partial/\partial t$.
The TDLDA and its small amplitude limit has been shown to be a reliable tool
to calculate strong resonances in the optical response of metallic
clusters\cite{ya96,ru96} and molecules such as polyenes\cite{lu94}.
Our method to solve the equations is a very straightforward
one, but it deserves some discussion because there are many implementations
of small-amplitude TDLDA in the literature, and our method is not
conventional in condensed-matter or chemical physics. We use a Hamiltonian 
of the usual Kohn-Sham
form, taking a local-density approximation for the non-Coulombic
electron-electron interaction, and a pseudopotential to describe
the carbon ions.

However, the calculations with {\it ab initio} Hamiltonians are very 
time consuming, and it is important to analyze the results 
to gain understanding of the functional 
dependence of the excitation energy on the cluster properties.
This will be done in Sect. 3; our result briefly is that the
resonance frequency $\omega_n$ in chains scales with $n$ as
$\omega_n\propto \sqrt{\ln n}/n$.  
This applies to both chains and rings, with different coefficients
of proportionality.

Part of the interest in the specta of carbon chains comes from 
astrophysics, the
question of the composition of interstellar matter.  Absorption
bands are seen which may be due to carbon clusters, but specific
identification of species remains controversial\cite{fu93,wa94}.
The TDLDA is probably only reliable to 10\% or so on the frequencies
of the strong transitions, which is not enough to use the theory
to assign unknown transitions.  However, it might be that 
the theory could be used to predict the shifts due to small
changes in structure, for example by adding another atom at 
the end of the chain.  We plan in the future to investigate various molecules
that are similar to the carbon chains and see how large 
the perturbations are.
\section{TDLDA}
\subsection{General}
Time-dependent mean field theory is a powerful tool to
calculate the collective excitations of quantum many-particle systems,
and it has been widely applied in cluster physics. There are
a number of implementations of the theory, all starting from the time-dependent
Schroedinger equation.  With the local density approximation for the
electron-electron interactions, the equation has the form of a 
time-dependent Kohn-Sham equation,
\be
i \hbar \frac{\partial}{\partial t}\psi_i(\vec r,t)
= \left\{ \frac{\hbar^2}{2m}\nabla^2 
         + \sum_{a \in ion} V_{ion}(\vec r - \vec R_a)
         + e^2 \int d\vec r' \frac{\rho(\vec r',t)}{|\vec r - \vec r'|}
         + V_{xc}\bigl(\rho(\vec r',t)\bigr) \right\} \psi_i(\vec r,t).
\ee
Here $\psi_i$ represents a single-particle wave function, $n$ is
the electron density, $V_{ion}$ is the ionic core
potential, and $V_{xc}$ is the potential associated
with exchange and correlations,   For our purposes, $\psi_i$ is not 
perturbed greatly from the ground state wave function, and a small
amplitude approximation justifiable.  This leads to 
the RPA and the linear response method of solution.  We believe that
for large numbers of particles, the most efficient technique to treat
LDA Hamiltonians is the straightforward integration of the time-dependent
equations of motion.  The argument is given in the Appendix. 
The optical response of the light chains C$_{3,5}$ and
C$_7$ has been calculated with the configuration-interaction 
method of quantum chemistry\cite{pa88,ko95}, but this brute-force technique is 
impractical in large clusters.

Our numerical method is taken over from the nuclear physics\cite{Fl78}, 
where it is called the TDHF theory.  The efficiency of the method requires 
that the Hamiltonian matrix be sparse, and this is fulfilled by using a 
coordinate mesh to represent the wave function.  It is also necessary
for numerical efficiency to use Hamiltonians whose energy scales are
not too large; this means that one treats only valence electrons and
uses pseudopotentials for $V_{ion}$ to take into account the effects
of the core electrons.  Thus we treat dynamically only the four valence electrons
of carbon.  Our psuedopotentials are calculated according to the
commonly accepted prescriptions\cite{tr91,kl82}.  The exchange-correlation 
potential $V_{xc}$ is taken from ref. \cite{pe81}.  

Some details of our integration algorithm are given in the Appendix.  
The important numerical parameters are the spatial mesh size, 
$\Delta x$, the number of mesh points $M$, the time step $\Delta t$, and 
the total length of time integration $T$.  The values for these
parameters to C$_n$ clusters are discussed below.

\subsection{Initial conditions}
     We wish to start with the Kohn-Sham ground state wave function,
and in principle the geometry of the ions as well as the electron
wave functions are determined by minimizing the Hamiltonian function.
However, a full minimization of the structures is quite time consuming,
and we believe that small variations in bond lengths will
not have a significant effect on the collective excitations.
We therefore take geometries from outside rather
than calculating {\it ab initio}.
For simplicity, we
fixed the nearest neighbor distance of the carbon atoms at
1.28 \AA.  This is the average LDA equilibrium distance for large circular 
rings or long chains.

To calculate the response, our initial wave functions are perturbed from
the static solutions $\phi_i$ by a velocity field
\be
\psi_i(\vec r,0) = {\rm e}^{ikz}\phi_i(\vec r).
\ee
The momentum $k$ is set to be small to
ensure that the response is linear (we use typically $k=0.1~$\AA$^{-1}$).
The real-time evolution of the dipole moment is obtained as
\be
z(t) = \sum_i <\psi_i(t) | z | \psi_i(t)>,
\ee
and its Fourier transform in time gives the dipole strength
function $S(\omega)$
\be
S(\omega) = {2\omega m\over \pi k} \int_0^\infty \,dt\,z(t) \sin( \omega t).
\ee
The strength function defined in this way is related to the
oscillator strength $f$  by
\be
\int d \omega S(\omega)  = f.
\ee

\subsection{Results}
  We first discuss the numerical parameters needed for carbon
clusters.  We found that a mesh size of $\Delta x = 0.3$ \AA~is
needed to adequately represent the $sp$-orbitals of carbon.  This
is more than a factor of two finer than the mesh size parameter
needed for the $s$-orbitals of alkali metals.  The wave function
is represented on this grid within the interior of a cylindrical
volume.  The size of the cylinder is chosen to include all mesh
points within 4 \AA~of each carbon atom.  With this procedure,
the wave function for the C$_7$ cluster requires 30,000 mesh points.

   There are two numerical time parameters.  The first is the
time step in the integration.
In the carbon calculations we used 
$\Delta t = 0.001$ eV$^{-1}$, which is an order of magnitude smaller
than the value needed for alkali metal clusters.  This corresponds
to the order-of-magnitude difference between the maximum kinetic
energies in the meshes used for systems, 0.3~\AA~in carbon versus
0.8~\AA~in alkali metals. The other numerical time parameter is
the total integration time $T$.  The inverse of this sets the
scale for the energy resolution in the response; we use $T=10$ eV$^{-1}$
to make visible structure on the energy scale of 0.1 eV.

The response of
a typical case, C$_7$, is shown in Fig. 1.  Plotted is the
integrated oscillator strength as a function of excitation energy.  
The upper curve shows 
the response parallel to the chain, and the lower curve shows 
the perpendicular response. The total sum rule ($f=4n=28$, 
four valence electrons per carbon atom),
is satisfied
to within 10\% by the calculation\footnote{In principle, the TDLDA conserves
the sum rule exactly if the Hamiltonian is local.  This is not the
case when the core electrons are omitted, because the resulting
pseudopotentials have a nonlocal character\cite{kl82}.}
The longitudinal response has a strong excitation at 5.3 eV.
About 1/3 of the 28 units of oscillator strength is in this excitation. The
perpendicular response has no corresponding strong excitation
below 10 eV.   The small wiggles in the calculated response are artifacts 
of the truncation in
the Fourier transformation.  
  In our case here,
the effect is obviously small.  To understand better the nature of the
5.3 eV collective excitation, it is
helpful to know the characteristics of the orbitals near the Fermi
surface and the corresponding single-particle transitions.
The single-particle level scheme, giving by the eigenenergies of the
static Hamiltonian, are shown in Fig. 2.
The lowest strong single-particle transition is between the highest 
occupied and lowest empty $\pi$ orbitals.  This gap is 2.1 eV, roughly
half the frequency of the TDLDA mode.  Note that there are two $\sigma$
orbitals within the $\pi-\pi^*$ gap, but these electrons are localized 
on the outer ends of the carbon chain and do not couple strongly 
to the other states. 

The 2.1 eV $\pi-\pi^*$ transition has a very large oscillator
strength ($f=10.2$), which equals the strength of the TDLDA mode to 
within 4\%.  The magnitude of $f$ may be understood qualitatively 
as equal to the number of $\pi$ electrons. To a good approximation,
the longitudinal dipole operator on $\pi$ electrons in a chain
keeps them in the $\pi$ manifold, so the
sum rule is approximately conserved within the $\pi$ manifold.  The 
number of them in an odd-$n$ chain is $N=2n-2$, filling the
molecular orbitals in the usual way.  Thus, if the lowest transition 
would exhaust the sum rule, the oscillator strength would be
$f=12$ for the twelve $\pi$ electrons in $C_7$.  This is to be
compared with $\sim 10$ from the LDA wave functions.

We have calculated the excitation energies and oscillator strengths
for the collective transition for carbon chains in the 
range\footnote{In the ground state of the  even-$n$ clusters with 
chain geometry, the highest occupied orbital is doubly degenerate and 
we assumed each spatial orbital to be occupied
by one electron to make a triplet ground state.
For other cases,
all the occupied orbitals are completely filled.}
 $N=3-20$, 
and the results are tabulated in Table I. When more than one transition is seen,
the average excitation energy and the summed oscillator strength are given.
The features described for
$C_7$ apply systematically these chains. The oscillator
strength is roughly given by the number of $\pi$ electrons; the frequency
of the TDLDA mode is larger than the unperturbed excitation energy by
a factor that ranges from 2 in the light chains to 3 in the heavier ones.
The frequencies from Table I are graphed in Fig. 3 and one can see
that they vary smoothly with chain length.  
The filled circles are experimental points for
odd $n$ from ref. \cite{ch82} and \cite{fo96}.  We see rather good agreement
for the larger chains. The clusters  C$_{3,5,7}$ have
been calculated by the CI method of quantum chemistry\cite{ko95}, and the
predicted strong transition is in the same region, at 8.10, 6.35, and 5.54 eV,
respectively, which are close to our results of 8.1, 6.4, and 5.3 eV. 
The oscillator strength for C$_3$ is calculated in ref. \cite{ko95} 
as about 1.1. The oscillator strength given in Table I is 3.1 for C$_3$,
but this is the strength along the chain direction. Dividing the strength 
by three to average over spatial orientations gives f=1.0, in 
agreement with the quantum chemistry calculation.
While there is experimental data for the excitation energies of odd-$n$ 
chains in the range $n=7-15$, there is no corresponding data on the even 
chains.  We are confident that the systematics is
smooth going over odd and even $n$, so it should be possible to observe
the mode in even-$n$.

  Ref. \cite{fo96} also reported weaker transitions in odd-$n$ chains
at lower energy. The authors in
ref. \cite{fr95,fo95} observe transitions in heavier even-$n$ clusters which
fall within the same systematics as the weak odd-$n$ transitions.
We believe that the even-$n$ transitions may be associated with 
the single-particle transition across the energy gap.  There are
two inequivalent electrons at the Fermi surface due to the partial
orbital occupancy, and one combination becomes the high-frequency
collective state and the other combination remains near the gap
energy with only a small oscillator strength.  In Fig. 4
we show
the systematics of the low transition, compared to the calculated
$\pi-\pi^*$  gap energy.  

%The single-particle response, or free response, of C$_{10}$ is shown 
%in Fig. 3.  There are
%two transitions at about 2 eV, roughly half the frequency of the 
%collective excitation.

  We have also calculated the in-plane response of ring configurations
in the range $N=7-15$. For simplicity we assumed uniform circular rings,
although the structures may have some distortion\cite{ra87}.  The results 
for the collective $\pi$ transition are shown in Table II.  The ratio of
collective frequencies for rings to chains is plotted in Fig. 5.
We see that the rings are predicted to have a collective frequency about twice that
of the chains with the same number of atoms.

\section{Interpretation}
    Our goal in this section is to characterize the $n$-dependence of the
excitations, and further the dependence on shape, whether chain
or ring.  It is well known that the Fermi gap (or HOMO-LUMO gap) 
$\Delta \epsilon$ in a linear
chain depends on $n$ as 
\be
\Delta \epsilon \propto {1\over n}.
\ee
This function fits well the systematics of our calculated single-particle
transition energy, but 
it does not describe the collective excitation.  This may be
seen from the log-log plot of the energies in Fig. 6.  The figure shows
that electron-electron interaction plays an important
role not only in the absolute frequency of the transition but
in its functional form as well. The inadequacy of eq. (6) to describe
the systematics of the strong excitation was also mentioned in 
ref. \cite{fo96}.

In purely classical physics, the analogous problem is the
plasma resonance in a needle-shaped conductor.  The closest
problem that can be treated analytically is plasma resonance in
a ellipsoidal conductor.  Ref. \cite{bo83} derives the formula,
\be
\label{huff}
\omega^2 = {1-e^2 \over e^2}\Big(-1+{1\over2e}\log{1+e\over
1-e}\Big)\omega_0^2
\ee
where $e$ is related to the ratio of short to long axes,
$R_\perp/R_\parallel$, by  
\be
e^2 = 1 - \Big( {R_\perp\over R_{\parallel}}\Big)^2.
\ee
This formula has been applied to the collective $\pi$ excitations
in football-shaped fullerenes \cite{br96}.  For our problem,
we obtain the fit shown in Fig. 6, treating $\epsilon$ and
$\omega_0$ as adjustable parameters.  It describes the systematics
very well.  Of course, the carbon chain is not ellipsoidal, and
we should seek rather that analytic behavior of the resonance in
a conducting cylinder.

     The behavior of electromagnetic resonances on infinitely
long wires is known from classical electromagnetic theory.
The dispersion formula for the one-dimensional plasmon
on a long wire reduces to the following expression in the
long-wave length, thin wire limit\cite{go90,li91}. 
\be
\omega^2 = { 4 \pi n_e e^2\over m} q^2 \log{1\over qa}
\ee
where $q$ is the reduced wave number of the plasmon,
$n_e$ is the density of electrons per unit length, and
$a$ is the radius of the wire.  For a finite wire, 
the lowest mode would have a $q$ varying inversely with
the length of the wire $L$.  Thus the lowest mode would
behave as
\be
\label{logn/n}
\omega\propto {\sqrt{\ln(L)}\over L}.
\ee
This in fact is the asymptotic behavior of eq. (\ref{huff})
in the limit of large $R_\parallel=L$.
Taking $L\propto n$, the frequency dependence in chains would be
\be
\label{logn/n2}
\omega\propto {\sqrt{\ln(n)}\over n}
\ee
This behavior can be extracted from a more quantum approach 
using the polarizability estimate of the collective 
frequency\cite{de93},
\begin{equation}
\label{alpha-estimate}
\omega^2 = {\hbar^2 e^2 N \over m  \alpha}.
\end{equation}
Here $N$ is the number of active electrons and $\alpha $ is
the polarizability.  This formula is derived from the ratio of
sum rules, and $N$ may be identified with the oscillator strength
$f$ associated with the transition.  We established in Sect. 2
that the oscillator strength in the $\pi$ manifold of states
is given roughly by the number of $\pi$ electrons, and scales
accordingly with $n$.

We next estimate the polarizability. The asymptotic behavior for 
large $n$ can be determined under the assumption that the chain behaves
as a perfect conductor. 
The electrons in a perfect conductor will respond to an external
field to restore the internal electric field to zero.
This gives an implicit equation for the
electron density in terms of the external electric field ${\cal E}$,
\be
z{\cal E} = e\int dz' \delta  n(z') V(z-z')
\ee
Here $\delta  n(z)$ is the induced linear electron density, obtained
by integrating the induced ordinary density over transverse coordinates.
If this is solved for $\delta n(z)$,
the polarizability is then obtained as the ratio of the induced
dipole moment to ${\cal E}$,
\be
\alpha = {1\over {\cal E}} \int dz' z'\delta n(z')
\ee
to determine asymptotic behavior, we note that $V(z)$ is strongly peaked at zero, 
so we may approximate it
as a $\delta$ function times $\int V dz$:
\be
\int dz' \delta  n(z') V(z-z')\approx \delta  n(z) \int dz' V(z')
\ee
The $V(z)$ behaves as $1/|z-z'|$
at large separation, so integral on the right hand side depends
logarithmically on the integration limits.  Thus we may estimate
the integral as
\be
\int dz' V(z') \approx \int_{-L/2}^{L/2} dz' V(z') 
\approx 2 \log({L\over a}).
\ee
Here $a$ is a length having the order of magnitude of the transverse
dimension of the wire. We thus obtain
\be
\delta  n(z) \approx {z{\cal E}\over 2 \log(L/ a)}.
\ee 
{From} eq. (4), this implies that the polarizability is
\be
\alpha= {L^3\over 24 \log(L/ a)}
\ee
We now insert this in eq. (12) and use $n\sim L$ to obtain eq.
(\ref{logn/n2}). The fit with this function is shown in Fig. 7.  It obviously
describes the asymptotic behavior much better than eq. (6),
which is also shown in the figure.

We finally discuss the relative frequencies of the modes in 
chains and rings using eq. (12).  We first examine the oscillators
strengths, needed for the numerator of eq. (12). Naively we
would expect similar values, since the number of $\pi$ electrons
in a ring from orbital counting is given by $N=2n$.  However,
comparing Tables I and II, we may see that the in-plane ring values are 
only 2/3 the
chain values along the axes\footnote{The reason for the lower value is that
the electric field in the plane of the ring is partly transverse
and partly longitudinal with respect to the C-C bond axes.  This implies
dipole excitations lie partly outside the $\pi$ manifold of states.
Thus part of the sum rule is shifted to higher energy excitations.}.
Much more important for the frequency change is the differing 
polarizabilities of chains and rings.  We have made a similar asymptotic
estimate of the ring polarizability, and we find that it also
increases logarithmically with the circumference of the ring.  The
ratio of polarizabilties of a thin wire ring compared to a straight
wire of the same length is about a factor of 6.  The two factors of
the oscillator strength ratio and the polarizability ratio combine
in eq. (12) to produce a frequency shift by a factor of two.  This
is indeed what the TDLDA gives asymptotically (see Fig. 5), confirming
the polarizability and oscillator strength argument we made here.

\section{Conclusion}
  We have demonstrated that the collective $\pi$ transition in
carbon chains and rings behaves in a very systematic way, 
calculable to $\approx 10\%$ accuracy by TDLDA, and understandable
in macroscopic terms.  This should give one confidence in
using these transitions to infer the shape of the clusters.
We hope in addition that the smooth dependence of the transition
would allow its perturbation going between similar structures
to be calculable to accuracy of interest for spectroscopic identification
purposes.

We thank R.A. Broglia for calling our attention to \eq{huff}. This work
is supported by the Department of Energy under
Grant No.~DE-FG06-90ER40561, and by a Grant in Aid
for Scientific Research (No. 08740197) of the Ministry of
Education, Science and Culture (Japan).  Numerical calculations
were performed on the FACOM VPP-500 supercomputer in RIKEN and
the Institute for Solid State Physics, University of Tokyo.
 
\appendix
\section*{Numerical aspects}
    The method we use, explicit time integration of eq. (1) in
the coordinate space representation, can be shown to be the
most efficient of the methods in use for large systems when there
are no symmetries to reduce the dimensionality.  The
coordinate space representation is also the most efficient for
the static problem under similar conditions\cite{ma89}.  

Let us first consider the matrix RPA in a particle-hole
configuration space representation.  This method has been applied to
sodium clusters in ref. \cite{ko91} and to 
C$_{60}$ in ref. \cite{ya93}.  The number of numerical operations
to extract the eigenmodes and eigenfrequencies scales with the
dimensionality of the matrix $M$ as $M^3$.  The dimensionality
of the matrix is given by the number of particle-hole configurations.
A complete calculation, guaranteed to respect the sum rules and
conservation laws, requires that all occupied and unoccupied orbitals
be included in the space.  The single-particle space will have a
dimensionality that is proportional to the size of the system,
and the number of unoccupied orbitals will thus be proportional
to the number of particles $N$.  Thus the dimensionality with this
method scales with $N$ as $N^6$.  Note that this could be much reduced
by truncating the particle-hole basis.  For example, in our study of
collective $\pi$ transition in carbons chains, the orbitals away from the
Fermi energy are not so important. However, we do not know a systematic
scheme for truncation that would preserve the oscillator strength and
lead to a more favorable $N$ dependence for the algorithm.

Another widely used technique is the linear response method. This has
been applied to the molecule N$_2$ in ref. \cite{ja96} and to alkali metal
clusters in ref. \cite{ru96}.  Here the object one
calculates is the density-density response function.  In a coordinate
space representation, its dimensionality is the number of points in
the coordinate space mesh.  Thus the dimensionality of the matrix scales as
$M \propto N$.  The matrix operation required in this case is inversion
rather than diagonalization, but it also goes as the cube of the dimension.
So this method scales as $O(N^3)$.  Note that we could have used
another representation of the response, such as momentum space,
and still obtained $O(N^3)$ scaling.

The advantage of direct coordinate space methods is that one can take advantage
of the sparse character of the Hamiltonian in that representation.  The
Hamiltonian has a dimensionality that scales $M\propto N$, and the number
of operations required to apply the Hamiltonian to a single-particle wave
function is also $O(N)$ because of the sparseness.  Performing the 
operation on all $N$ particles is then $O(N^2)$, and this gives the
scaling for the method.  Note that there is a large prefactor, because
the equations have to be integrated many time steps.  However, the
time integration depends only on the energy scales in the Hamiltonian
which does not change
with $N$.   Finally, we note that the advantage of this method would 
apply to any representation of the wave function that allows a sparse 
Hamiltonian matrix.

For the numerical aspects of integrating the Kohn-Sham 
equation, we follow closely the method used in ref. \cite{Fl78} to
solve the time-dependence mean field equations in nuclear physics.
The algorithm must insure energy conservation and norm conservation
to very high accuracy to be useful.  In addition, for the results to
be converged, the time step of the integration must be small compared
to the inverse frequencies of the density oscillations.  
Fixing the time step by this criterion, 
energy conservation is achieved by using an implicit
method to integrate the equations.  One can show that the integration
over the time step will conserve energy if the mean field is computed
with a fixed density given by the average of the initial and final
densities, for Hamiltonians with ordinary two-body 
interactions\footnote{A more general condition, valid for nonlinear
density dependencies such as in the correlation-exchange potential,
is given by taking the potential in the single-particle Hamiltonian
as
$$
V(r) = {v[n_+(r)] - v[n_-(r)]\over n_+ - n_-}
$$  
where $v$ is the potential energy functional of density and 
$n_\pm$ are the densities at the beginning and end of the
time step.}.
In our computer program, we obtain sufficient accuracy for our time
steps by using a simple predictor corrector method to find the
average density and integrate over the time step.

We have now reduced the problem to the time integration of a fixed
Hamiltonian.  Since the Schroedinger equation is linear, we can integrate
straightforwardly by using the Taylor series expansion
of the time evolution operator,
\be
\phi(t+\Delta t) = e^{-i H \Delta t} \phi(t) = \sum_{n=0}^{k_{max}}
{1\over n!}(i H
\Delta t)^n \phi(t)
\ee
We found that the fourth-order approximant, $k_{max}=4$, 
is sufficient with our time steps to insure norm conservation to the 
needed accuracy.

\begin{table}
\caption{Excitation energy and strength of carbon chains 
in TDLDA}

\begin{tabular}{rrrrr} 
Size & $E_{\rm free}$ [eV] & $f_{\rm free}$ & 
       $E_{\rm TDLDA} [eV] $ & $f_{\rm TDLDA}$ \\
\hline
  3 &  4.15   &   3.5   &    8.1    &   3.1  \\
  4 &  3.89   &   5.2   &    7.2    &   4.5  \\
  5 &  2.81   &   6.9   &    6.4    &   6.3  \\
  6 &  2.72   &   8.6   &    5.9    &   8.0  \\
  7 &  2.13   &  10.2   &    5.3    &   9.8  \\
  8 &  2.10   &  11.9   &    5.0    &  11.4  \\
  9 &  1.71   &  13.5   &    4.6    &  13.1  \\
 10 &  1.71   &  15.2   &    4.4    &  14.8  \\
 11 &  1.43   &  16.7   &    4.1    &  16.4  \\
 12 &  1.44   &  18.4   &    3.9    &  18.1  \\
 13 &  1.23   &  19.8   &    3.7    &  19.7  \\
 14 &  1.26   &  21.6   &    3.5    &  21.3  \\
 15 &  1.08   &  22.9   &    3.3    &  22.9  \\
 20 &  0.90   &  31.7   &    2.7    &  30.8  \\
\end{tabular}
\end{table}
\begin{table}
\caption{Excitation energy and strength of carbon rings 
in TDLDA}

\begin{tabular}{rrrrr} 
Size & $E_{\rm free}$ [eV] & $f_{\rm free}$ &
       $E_{\rm TDLDA}$ [eV] & $f_{\rm TDLDA}$ \\
\hline
  7 &  4.9   &  5.8    &    7.8    &   3.1  \\
  8 &  4.8   &  8.6    &    8.2    &   4.6  \\
  9 &  4.1   &  9.0    &    7.6    &   5.5  \\
 10 &  3.7   &  9.8    &    7.2    &   6.9  \\
 11 &  3.5   & 10.7    &    7.0    &   7.9  \\
 12 &  3.3   & 11.8    &    6.9    &   9.2  \\
 13 &  3.0   & 12.5    &    6.6    &  10.3  \\
 14 &  2.8   & 13.5    &    6.3    &  11.3  \\
 15 &  2.6   & 14.4    &    6.1    &  12.4  \\
 20 &  2.0   & 18.7    &    5.2    &  17.4  \\

\end{tabular}
\end{table}

\newpage

Figure Captions
 
Fig. 1. Optical response of the linear chain C$_7$, calculated in TDLDA.
The mesh spacing is $\Delta x = 0.3$ \AA~and $\Delta t = 0.001$ eV$^{-1}$.
The spatial grid has the shape of a cylinder, with 30,000 mesh points. 
The integration time was $10$ eV$^{-1}$.  \\
Fig. 2. Molecular orbital diagram for the C$_7$ chain in the LDA. On the
left are shown the lowest $\pi$ states, which correspond to the 7
orbitals of the tight-binding approximation.  On the right are the
highest occupied and lowest unoccupied $\sigma$ orbitals.\\ 
Fig. 3.  Systematics of the longitudinal mode in C$_N$ linear chains.
Circles: experimental, from ref. \protect\cite{fo96} and \protect\cite{ch82};
squares, TDLDA; triangles, quantum chemistry calculations from ref.
\protect\cite{pa88} and \protect\cite{ko95}.\\
Fig. 4.  Systematics of $\pi-\pi^*$ gap energies compared with
observed transitions from ref.
\protect\cite{fr95,fo95}.\\
Fig. 5. Ratio of the collective excitation energy in rings with 
compared to chains, from the TDLDA.\\
Fig. 6.  Systematics of the single-electron transition and the collective
excitation in chains, compared with functional fits of the form eq. (6) 
and eq. (7).\\
Fig. 7.  Comparison of the TDLDA collective excitation in chains
with eq. (6) and (11).\\
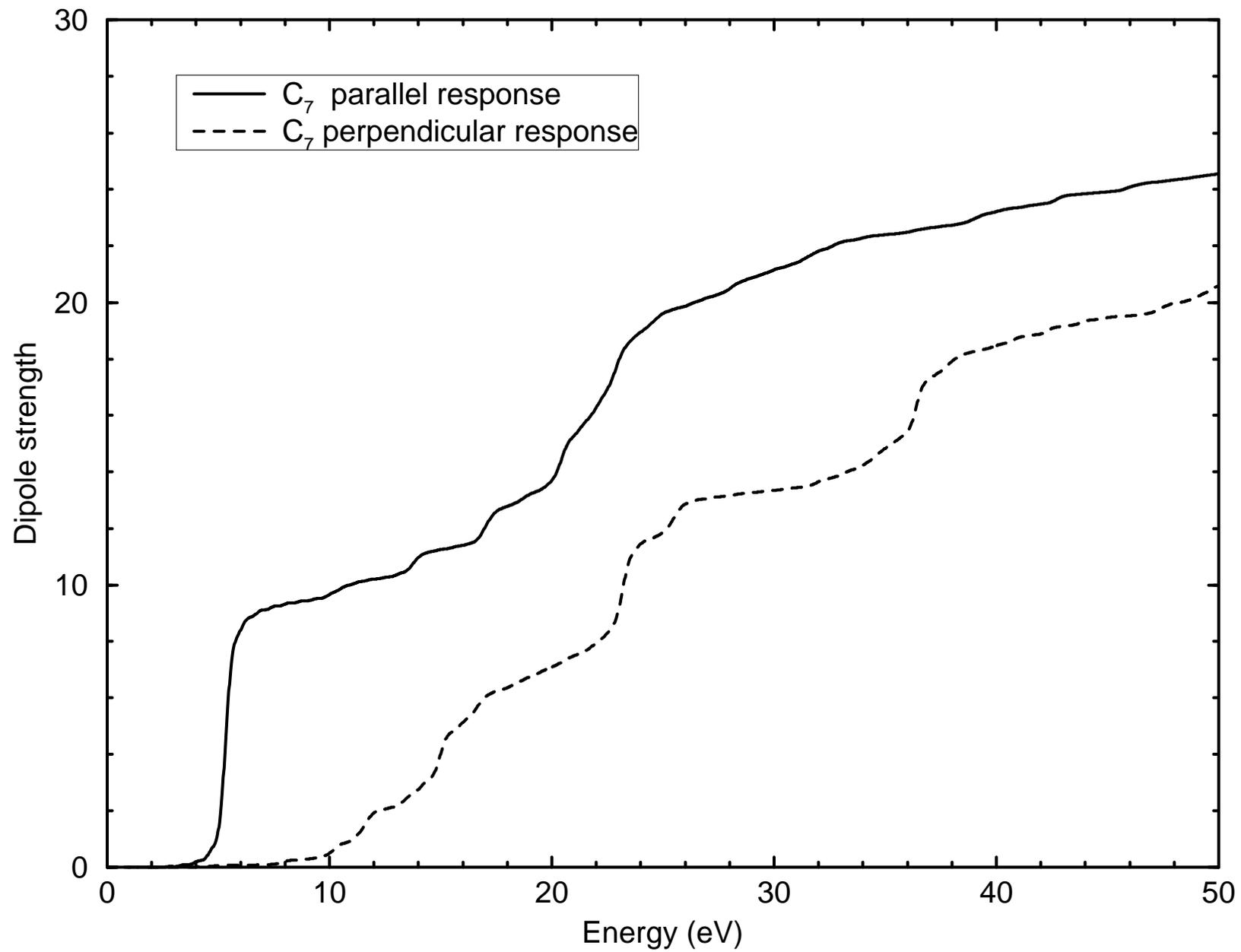
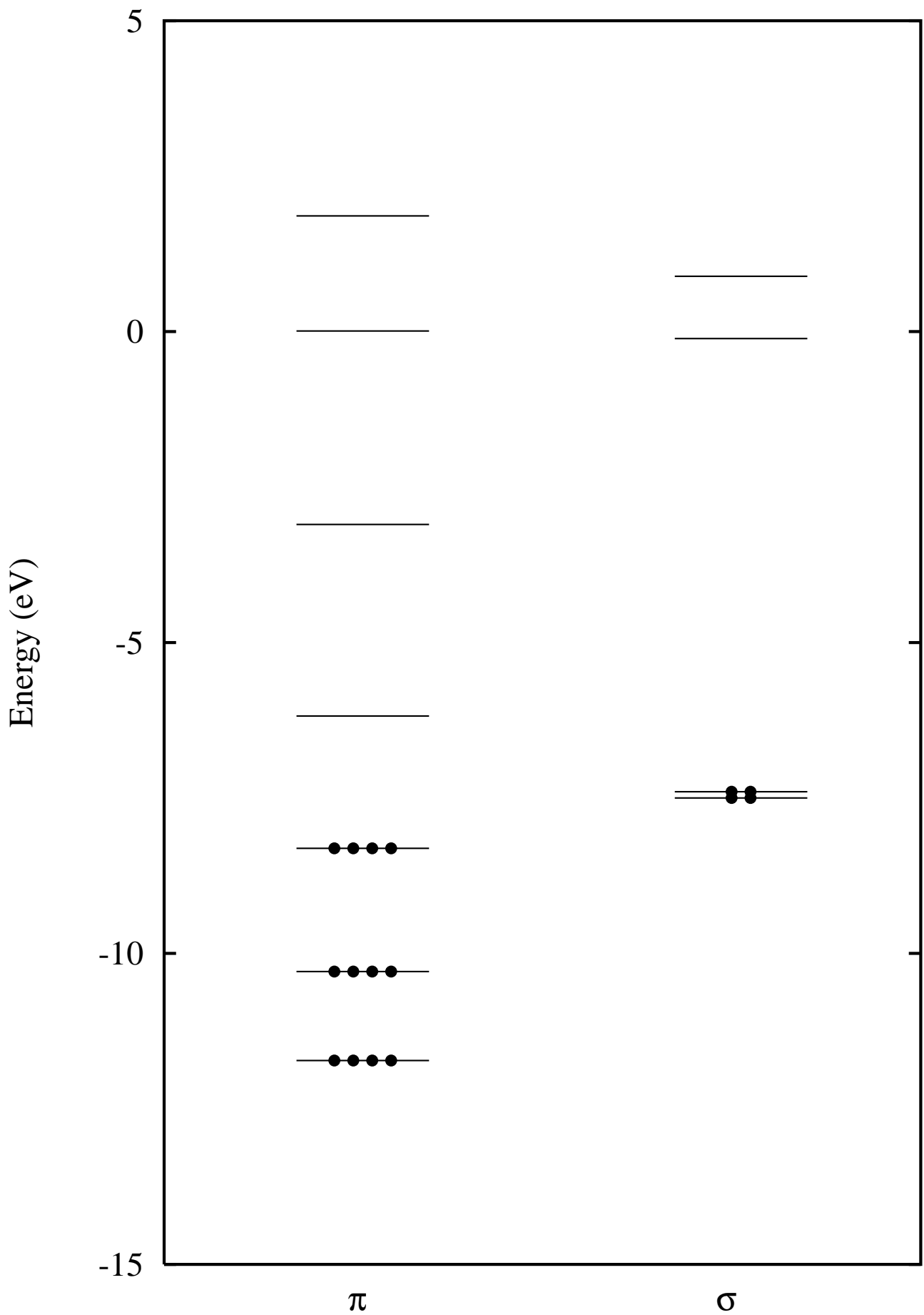
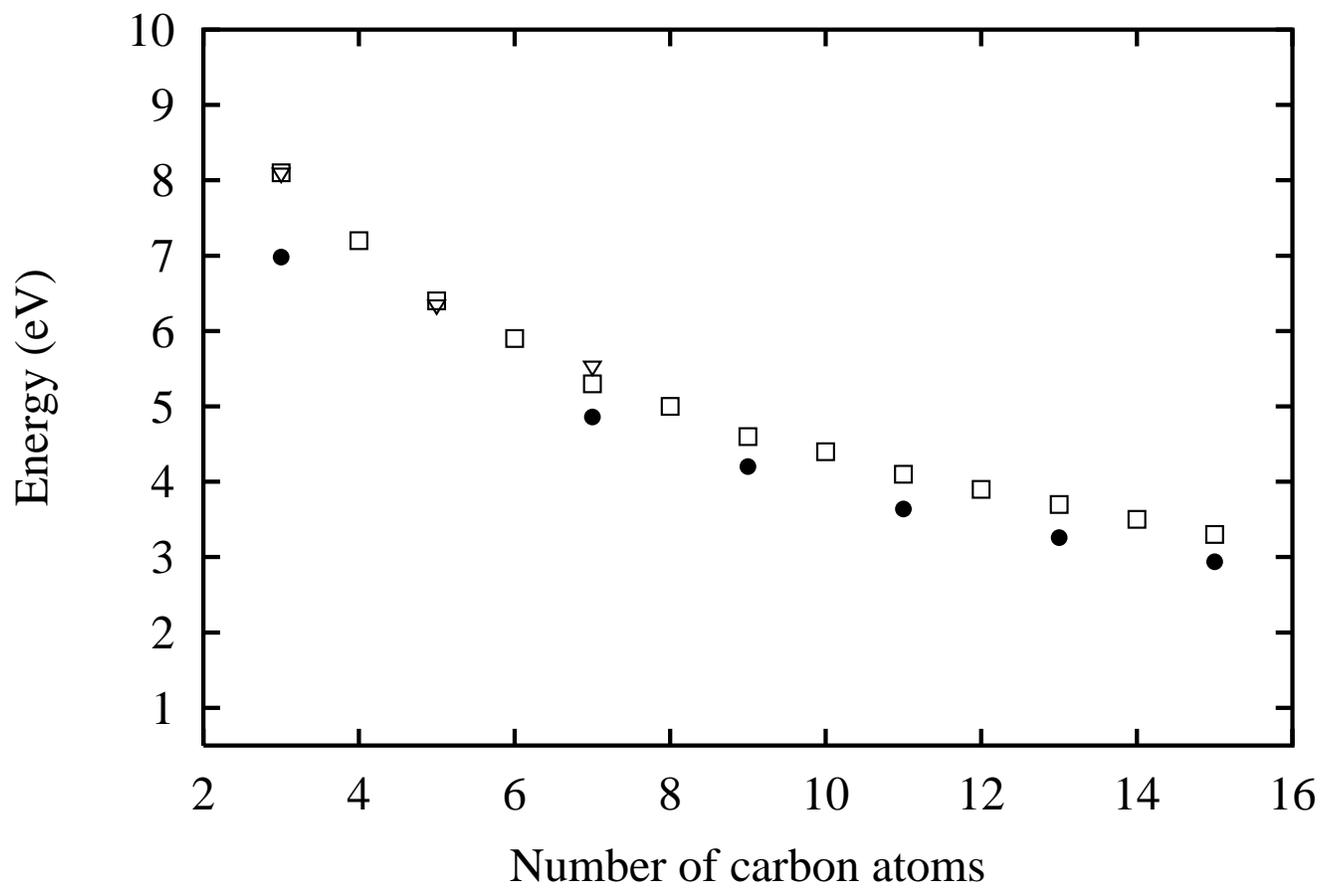
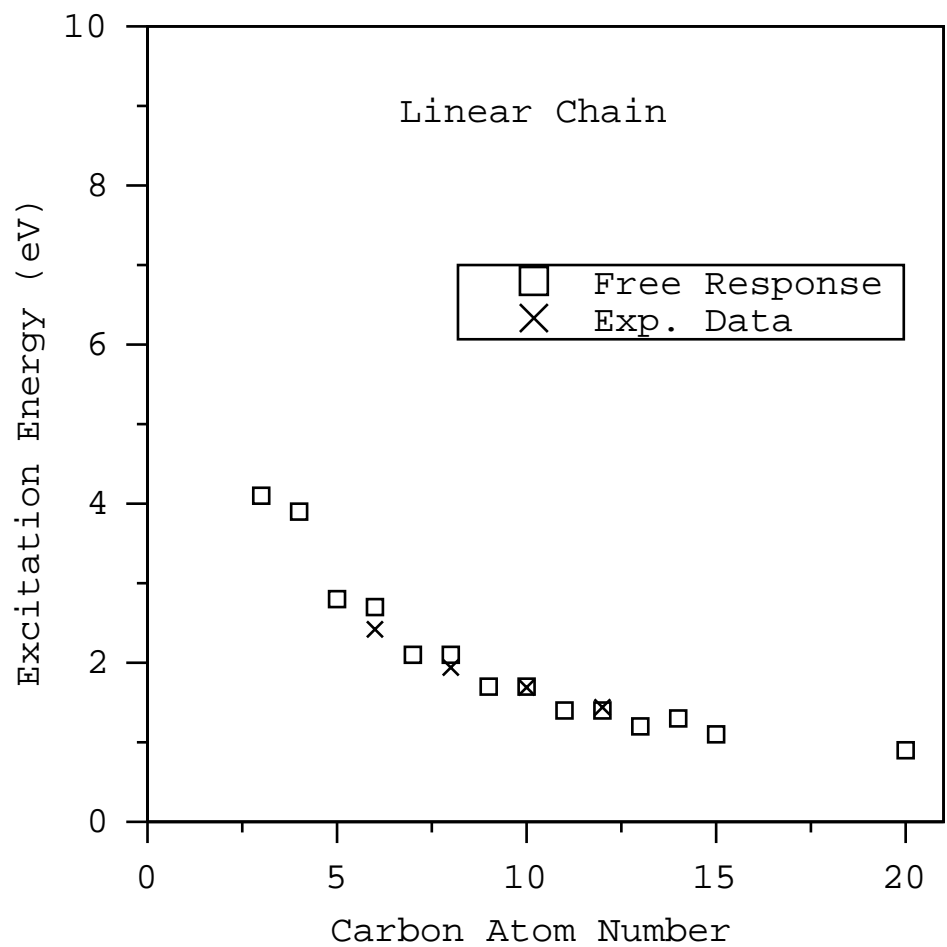
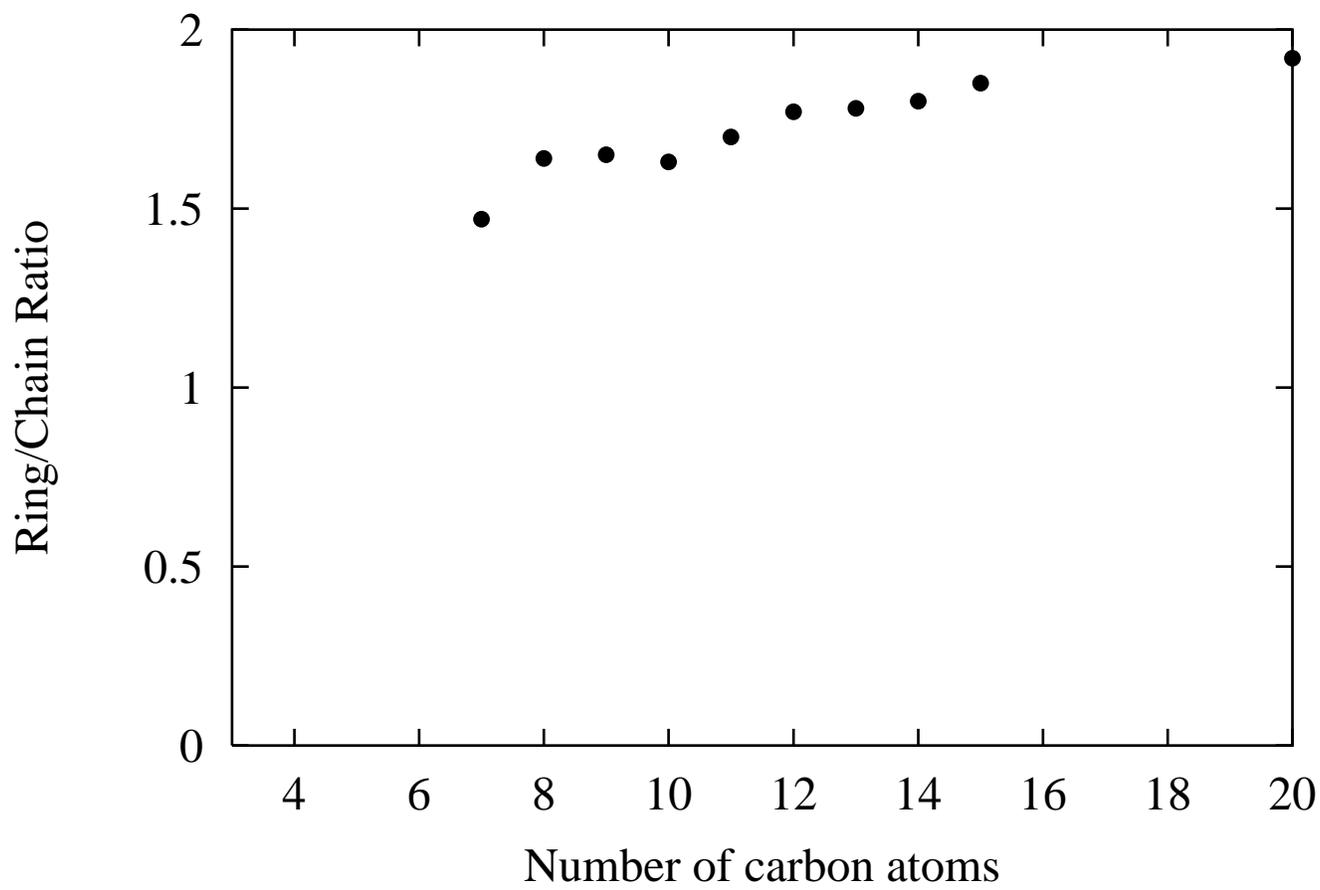
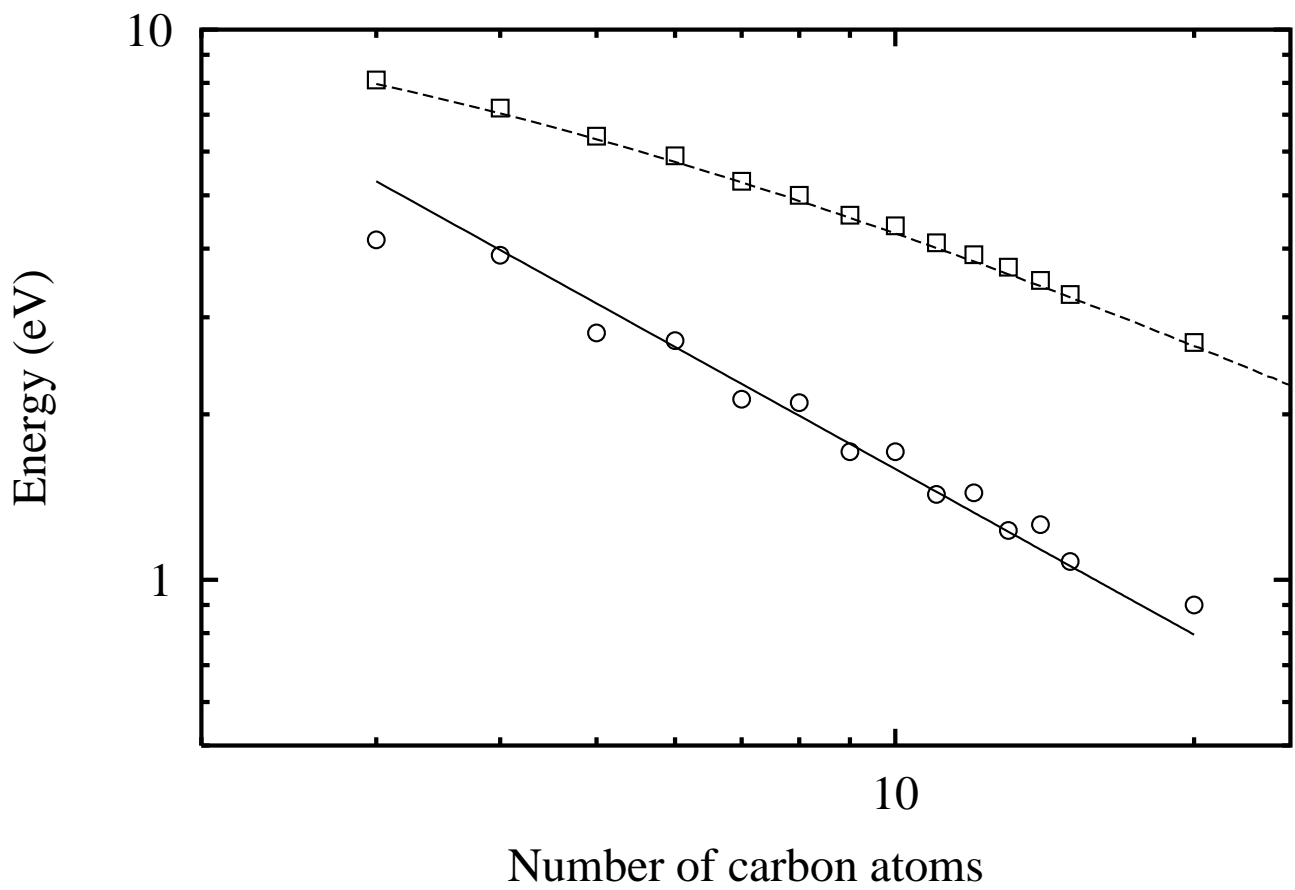
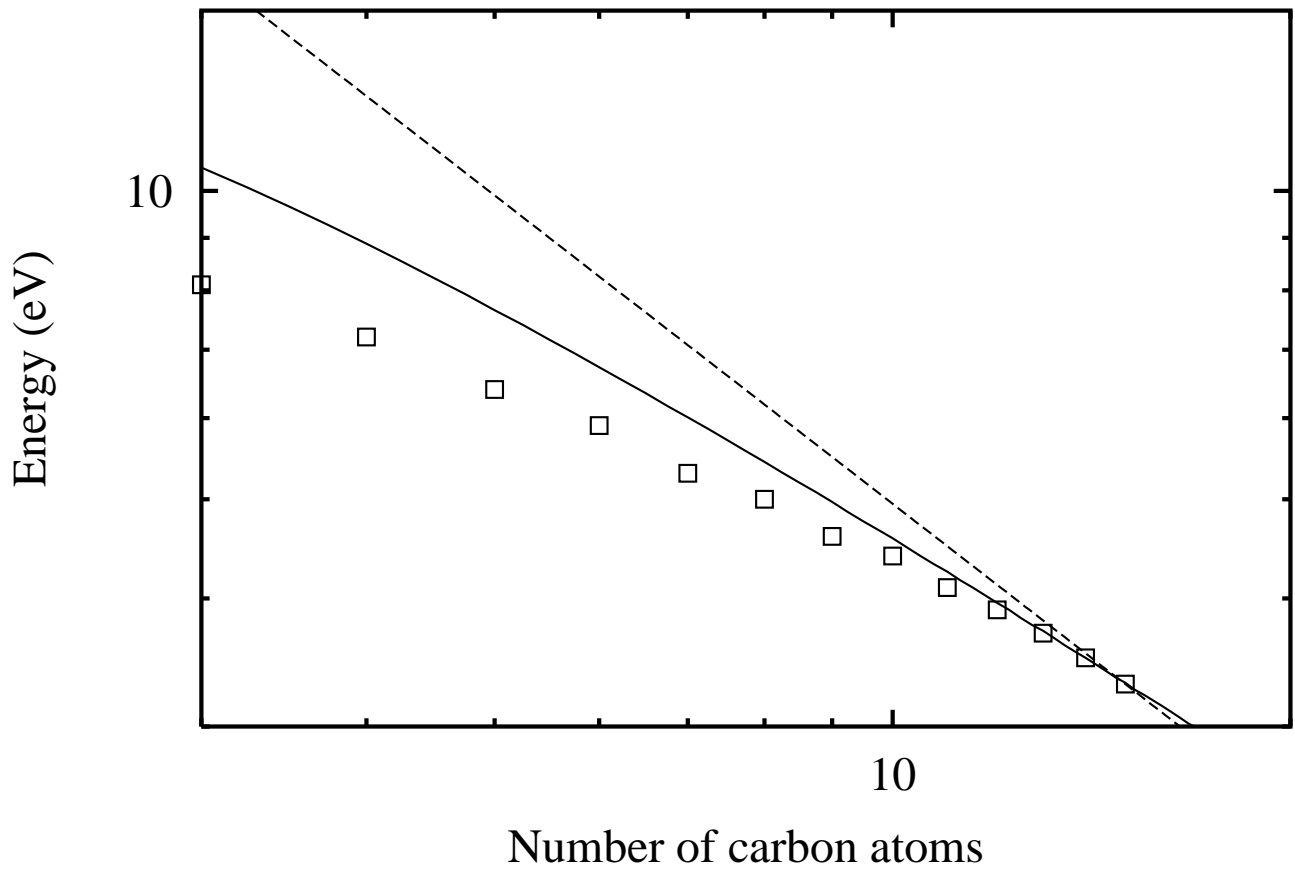
\end{document}